\documentclass[11pt]{article}
\pdfoutput=1
\usepackage{jheppub}
\usepackage{amsfonts}
\usepackage{amsmath}
\usepackage{mathtools}
\allowdisplaybreaks[4]         
\usepackage{amssymb}   
\usepackage{euscript}       
\usepackage[dvipsnames]{xcolor}
\usepackage{IEEEtrantools}
\usepackage{upgreek}
\usepackage{graphicx}
\usepackage{caption}
\usepackage{subcaption}
\usepackage[export]{adjustbox}
\usepackage[makeroom]{cancel}
\usepackage{physics}
\usepackage{tensor}
\usepackage{braket}
\usepackage{float}
\usepackage{color}
\usepackage[normalem]{ulem}
\usepackage{starfont}

\usepackage{hyperref}
\hypersetup{
	colorlinks =blue,
	linkcolor =blue,
	pdftitle={The quantum non-linear $\sigma-$ model RG flow and integrability in wormhole geometries},
	citecolor=MidnightBlue,
	filecolor=MidnightBlue,
	urlcolor=MidnightBlue,
}

\newcommand{\arxiv}[1]{{\tt
\href{http://www.arXiv.org/abs/#1}{#1}}}
\newcommand{\doi}[1]{{\tt DOI:\href{http://dx.doi.org/#1}{#1}}}

\makeatletter

\title{\huge The quantum non-linear $\sigma$-model RG flow and integrability in wormhole geometries}

\author[\text{a}]{Oscar Lasso Andino,} 
\author[\text{b}]{Christian L. V\'{a}sconez,} 

\affiliation[\text{a}]{Escuela de Ciencias F\'{i}sicas y Matem\'{a}ticas, Universidad de Las Am\'{e}ricas,\\
Redondel del ciclista, antigua v\'{i}a a Nayon, 170124, Quito, Ecuador}
\affiliation[\text{b}]{Departamento de F\'{i}sica, Escuela Polit\'{e}cnica Nacional,\\
Ladr\'{o}n de Guevara E11-253, 170525, Quito, Ecuador}

\vspace{0.3cm}
\emailAdd{oscar.lasso@udla.edu.ec, christian.vasconez@epn.edu.ec} 

\abstract{The target space of the non-linear $\sigma$-model is a Riemannian manifold. Although it can be any Riemannian metric, there are certain physically interesting geometries which are worth to study. Here, we numerically evolve the time-symmetric foliations of a family of spherically symmetric asymptotically flat wormholes under the $1$-loop renormalization group flow of the non-linear $\sigma$-model, the Ricci flow, and under the $2$-loop approximation, RG-2 flow. We rely over some theorems adapted from the compact case for studying the evolution of different wormhole types, specially those with high curvature zones. Some metrics expand and others contract at the beginning of the flow, however, all metrics pinch-off at a certain time. This is related with the fact that the flow does not converge to a fixed point when its starting geometry is the spatial sections of a Morris-Thorne wormhole, and therefore the corresponding non-linear $\sigma$-model is non-integrable/renormalizable. We present a numerical study of the evolution of wormhole singularities in three dimensions extending the theoretical estimations. Finally, we compute the evolution of the Hamilton's entropy and the Brown-York energy. }

\begin{document}

\maketitle

\newpage
\pagestyle{plain}



\section{Introduction}

Integrable string theories have been used for a better understanding of the holographic duality, specially in the case of $AdS_5 \times S^5$. However, the study of integrable theories goes beyond the AdS/CFT duality. The problem of classifying backgrounds  in string theory is a highly non trivial task. By knowing which string theories are integrable, we can know which are renormalizable, i.e. solved exactly without the use of perturbation theory. Thus, integrable theories can be studied by analyzing the quantum renormalization group flow. In particular, integrable $\sigma$-models are stable under the quantum RG-flow \cite{Fateev:2018yos}, thus, the non-linear $\sigma$-model RG-flow stable solutions will correspond to integrable theories which are renormalizable theories. Although this statement remains as a conjecture, it is strongly believed that it holds true \cite{Levine:2021mxi}.

The motion of a $2-$dimensional bosonic string is described by the non-linear $\sigma-$model 
\begin{equation}
S=\frac{1}{4\pi\alpha}\int \gamma^{\alpha\beta}\partial_{\alpha}x^{a}\partial_{\beta}x^{b}g_{ab}dV(\gamma),
\end{equation} 
were $\alpha>0$ is the string coupling constant. In order to quantize this model we have to introduce a momentum cut-off $\Lambda$, giving a family of quantum field theories parametrized by $\Lambda$. The matrix $\gamma$ is the metric of a $2$-dimensional Riemannian  manifold, called the world-sheet of the string.  $g_{ab}$ is the metric of a $n$-dimensional Riemannian manifold called the target manifold. Then, the renormalization group flow of the quantum non-linear $\sigma-$model is an evolution equation for the metric of the target manifold.

Given a Riemannian manifold $M^n$, whose metric $g_{ab}(\lambda)$ is parametrized by an affine parameter, we define an intrinsic geometric flow as an evolution equation of this metric. The evolution generates a family of metrics $g_ab(\lambda)$ through the equation
\begin{equation}
\frac{\partial g_{ab}}{\partial \lambda} = \beta_{ab}(g(\lambda));\;\;\;\; g_{ab}(0) = \tilde{g}_{ab},
\end{equation}
where $\beta_{ab}$ is a tensor built by $g_{ab}(\lambda)$, together with its first and second derivatives. In particular, when $\beta_{ab}(g(\lambda)) = -2R_{ab}(g(\lambda))$, we obtain the very well known Ricci flow. This flow develops singularities, and there has been a long debate about how to treat them\cite{Rflow:1}.\\
The metric of the target manifold will evolve with the energy scale and -in principle- this target manifold can have any geometry. In this article, we will focus on wormhole geometries. Our motivation is based on recent reports that point out that the Riemannian part of certain family of wormholes are stable under the Ricci flow \cite{Husain:2008rg}. We revisit the latter problem and try to answer some questions related to the wormhole stability. We are also interested in studying how this manifold will develop singularities.
In a geometric theory, such as the non-linear sigma model, the singularities will arise when the curvature of the target manifold goes to infinity. When the target spacetime is a wormhole, the singularities appear at the throat \cite{Husain:2008rg}. If these results hold true for all scale energies, we would have found an integrable/renormalizable theory. We will show that this is not the case. Although there is a seemingly stable behavior of the flow, when evolved to larger ``times'' it will develop singularities. For a better treatment of higher curvature zones, we make evolve the wormhole geometries under the RG-2 flow, the second loop quantum renormalization group flow of the non linear $\sigma-$model.\\

Wormhole spacetimes arise as solutions of the Einstein equations. Nowadays, they are at the center of the debate in the high energy physics community. They seem to be very useful for studying theoretical aspects of holography, e.g. \cite{Maldacena:2017axo}. In particular, we want to know how asymptotically flat wormholes evolve under Ricci flow, and under its higher curvature correction: the RG-2 flow. We show that the flow develops singularities in finite time when evolving a spherically symmetric asymptotically flat time symmetric foliation of a spacetime with a throat\footnote{The evolution of the metric of a Riemannian manifold is attached to the evolution of a Lorentzian manifold built with the original Riemannian metric, where the evolution is perfectly controlled.}.\\

Moreover, the Ricci flow have been used in the proof of the Thurston geometrization theorem \cite{,Perelman:2006un,Perelman:2006up}. These flows are usually very difficult to treat because of singularities. Therefore, it is interesting, from the mathematical point of view, to investigate how different geometries behave -and develop singularities- under a given flow.\\

There are analytical and numerical results about the evolution of spacetimes (or its Riemannian foliations) under different flows, in different contexts\cite{Headrick:2006ti,LassoAndino:2018zhb}. In \cite{Garfinkle:2003an}, the authors presented a numerical study of singularities in compact surfaces by making a comparison between an evolution of a family of metrics in $S^3$, with a $S^2$ neck-pinching. They found a certain type of critical behavior near the singularity formation. This critical behavior depends on the amount of corseting. An apparent similar critical behavior has been found in other types of metrics\footnote{See \cite{Gundlach:2007gc} for the study of critical behavior in gravitational collapse.}. Thus, some kind of critical behavior is expected, under very particular circumstances, when evolving a metric under these flows. However, remembering that the Ricci flow is a first loop approximation, it is not enough when dealing with higher curvatures. In these cases, it is necessary to consider a higher curvature flow, as the RG-2 flow. We expect that criticality is going to be ``enhanced"\footnote{The enhancement here means that it will take more time, compared with the Ricci flow, to develop singularities. However, singularities will always arise at a long time.} in some way. In \cite{Husain:2008rg}, starting with a time-symmetric foliation of a wormhole metric, the authors reported a seemingly critical behavior, which is related to the type of wormhole\footnote{The shape function of the wormhole is parametrized by an exponent, which is associated with different types of wormholes. In particular, $\delta=2$ is the Ellis wormhole.} considered as initial metric. Using a more powerful and stable numerical method, we will refine these results. We will use analytic results for the evolution of asymptotically flat spacetimes\cite{Oliynyk:2006nr}. See the results in two \cite{Solodukhin:2006ic} and three \cite{LassoAndino:2019lsa} dimensions.\\
We evolve some wormhole geometries for larger times (compared with those presented in \cite{Husain:2008rg}), and for geometries with higher curvature points. We also discuss about evolution under higher loop flows. Our method can be extended to flows coupled to matter coming from the string theory such as the generalized Ricci flow\cite{Garcia-Fernandez:2020ope}. These results will help us to approach the problem thinking about extending the results to higher order flows. 
This article is structured as follows. In Section \ref{two} we make a brief review of the flows that we are going to use. In Section \ref{three}, we study analytically the evolution of a foliation of the Morris-Thorne wormhole under the Ricci flow, we describe the conditions needed for our numerical setup. In Section \ref{four}, we numerically compute the evolution of the spatial sections of the Morris-Thorne wormhole under the Ricci flow and the RG-2 flow. We continue this Section evolving different asymptotically flat geometries under the RG-2 flow. In Section \ref{five}, we discuss about the results and future work directions.

\section{Geometric flow equation}\label{two}
We take a wormhole spacetime in the wide sense, namely a spacetime that has a minimal surface \cite{Visser:1997yn}. The Ricci flow had been successfully used when studying the evolution of different physical quantities\cite{Woolgar:2007vz,Solodukhin:2006ic,LassoAndino:2018zhb}. Even when there is no mathematical formulation of the quantization of the non-linear sigma model, it is considered that the approximation with the Ricci flow works only when curvature is small. We want to explore what happens when higher curvature flows are taken into account. Certainly, far from the throat, we expect a similar behavior as the Ricci flow. However, when curvatures go higher, near the throat, we expect to see the influence of the second loop term.\\

The RG-2 flow \cite{Gimre:etal2} with the DeTurck term is given by
\begin{equation}\label{rg2flow}
\frac{\partial g_{ij}}{\partial \lambda}=-2R_{ij}-\frac{\alpha}{4}R_{i\alpha\beta\gamma}R_{j}^{\,\,\,\alpha\beta\gamma}+\nabla_{(i}V_{j)}.
\end{equation}
This flow is parabolic only in the zones where $1+\alpha K_{ab}>0$, where $K_{ab}$ is the sectional curvature of the manifold \cite{Gimre:2014jka,Oliynyk:2009rh}. When $\alpha=0$ (Ricci flow), the flow is weakly parabolic everywhere. We take a time-symmetric foliation of a spacetime, which is in practice tantamount to drop-out the time component of the metric, getting as a result a 3-dimensional Riemannian metric. Later, we will evolve this resulting Riemannian metric. The parameter $\lambda$ (not to be confused with physical time) is an affine parameter that labels a continuum family of Riemannian metrics, and the $V^i$ vector generates diffeomorphisms along the flow\footnote{The presence of this vector is a manifestation of the flow's anthropomorphism invariance. In consequence, we have a freedom when choosing $V^{i}$. This vector is going to be restricted by the flow and it will be different for every flow. When we add the  RG-2 term with $\alpha<1$ we want to study the solutions of the Ricci flow perturbatively. In this article we are not interested in studying the RG-2 flow as a general flow for any value of $\alpha$.  The $\alpha$ has to be bigger enough such that the RG-2 term is above the numeric noise, but small, the RG-2 term has to be in the perturbative regime.}.

\section{Evolution of the Morris-Thorne wormhole}\label{three}

We are ready for evolving a metric. We start by finding analytically the evolution of the time-symmetric foliation of the Morris-Thorne metric \cite{Morris:1988cz}:
\begin{equation}\label{morrisf}
ds^2=\frac{1}{\left(1-\frac{b(r,\lambda)}{r}\right)}dr^2+r^2(d\theta^2+\sin^2(\theta)d\phi^2),
\end{equation}
\noindent
where $b(r,\lambda)$ is the shape function. It satisfies that $b(r_{o},0)=r_o$, at the throat $r_{o}$. Replacing the latter equation in the system (\ref{rg2flow}), we can compute the components of the RG-2 flow equation:
\begin{eqnarray}
\frac{\partial b}{\partial \lambda}&=&\frac{2(b-rb')}{r(r-b)}\left(1-\frac{\alpha}{4}\frac{b-rb'}{4 r^3}\right);\label{RG21}\\
0&=&\left(\frac{b+rb'}{2r}\right)+\frac{\alpha}{2}\left(\frac{5b^2-2rb'b+r^2b'^2}{2r^4}\right).\label{RG22}
\end{eqnarray}
\noindent
The equation (\ref{RG21}) is an evolution equation for the shape function\footnote{In this section, we consider the RG-2 flow without the DeTurck term.} $b(r,\lambda)$. Meanwhile, equation (\ref{RG22}) is a restriction over  $b(r, \lambda)$.\\
When $\alpha=0$ the restriction equation (\ref{RG21}) can be solved, leading to $b(r,\lambda)=C_{r}(\lambda)/r$. It shows that the shape function will change when it is evolved under the flow. In particular, replacing $b(r,\lambda)$ in equation (\ref{RG22}), we obtain\footnote{We have assumed that $dr/d\lambda=0$.}: 
\begin{equation}
C_{r}(\lambda)=-r^2W\left[ -\frac{b_{o}^2}{r^2} \exp \left(\frac{4\lambda-b_{o}^2r}{r^3} \right) \right],
\end{equation} 
\noindent
where $W$ is the Lambert function\footnote{The Lambert function is defined as the inverse function of $f(W)=We^{W}$.}. After the evolution under the Ricci flow, the shape function starts changing, and therefore the throat will change its size. Moreover, the shape function far from the throat will not change showing that asymptotic flatness is maintained during evolution. It is very difficult to see how this change happens in these coordinates. This is the simplest solution of the system. If we want to go further, numerical methods are necessary. Moreover, when we consider the RG-2 flow, it is really difficult to find an analytic solution of the system (\ref{RG21})-(\ref{RG22}). In order to study what is really happening at the throat, we have to move to a more useful set of coordinates. We need a coordinate system that somehow helps us to see the evolution of the size of the throat directly.

\subsection{Initial wormhole metric}\label{initialc}
The metric (in its original form) does not let us see the evolution of the throat size. We take as the initial metric the spherically symmetric ansatz \cite{Husain:2008rg},
\begin{equation}\label{worm:1}
ds^2=e^{2\Psi(\lambda,\rho)}\left(d\rho^2+R^2(\lambda,\rho)(d\phi^2+\sin^2(\theta)d\phi^2)\right),
\end{equation}
\noindent
with a DeTurck vector defined -because of spherical symmetry- as:
\begin{equation}\label{dturk:1}
V^i=V(\lambda, \rho)\partial_{\rho}.
\end{equation}

We will take $\Psi(t,\rho)=0$. Therefore, the metric (\ref{worm:1}) is reduced to:
\begin{equation}\label{metric1}
ds^2=d\rho^2+R^2(\lambda,\rho)(d\theta^2+\sin^2(\theta)d\phi^2).
\end{equation}

Using the latter metric and the DeTurck vector defined in (\ref{dturk:1}), the RG-2 flow equations (\ref{rg2flow}) become:
\begin{eqnarray}
\frac{\partial R}{\partial \lambda}&=&\partial^2_{\rho} R+\frac{(\partial_{\rho}R)^2}{R}-\frac{1}{R}-\frac{\alpha}{8}\left[\frac{((\partial_{\rho}R)^2-1)^2}{R^3}+\frac{(\partial_{\rho}^2 R)^2}{R}\right]+V\partial_{\rho}R,\label{floweq1}\\
\partial_{\rho}V&=&-2\frac{\partial^2_{\rho} R}{R}-\frac{\alpha}{2}\frac{(\partial_{\rho}^2 R)^2}{R^2}.\label{floweq2}
\end{eqnarray}

We note that when $\alpha=0$ the Ricci flow, with its correspondent restriction equation, is recovered. In the next section, we will solve numerically the equations system (\ref{floweq1})-(\ref{floweq2}) considering it as a perturbation of the Ricci flow system. We will do this by choosing a small $\alpha$ value and studying how the Ricci flow solution deviates perturbatively. We start by solving the Ricci flow system. A first approach was made in \cite{Husain:2008rg}, where the authors reported that for a given family of wormholes there is a criticality (when $\delta=1.259$). This means that the proposed metric remains the same when it is evolved by the Ricci flow. Our results confirm this statement, but within a limited given zone. However, if the flow evolve further, the throat will develop a singularity. Therefore, the metric \eqref{metric1} does not constitute an integrable/renormalizable model for the range of $\delta$'s that have been evolved numerically in this work. It is worth to note that in \cite{Husain:2008rg}, combinations of pure finite-difference schemes are employed, while we will use a pseudo-spectral method (appropriate for studying viscous fluids) that allows us to comparing the evolution of different initial metrics under the Ricci and the RG-2 flows, respectively. In order to proceed with the computations, we set initial and boundary conditions.\\
The asymptotic flatness in both directions is translated to the Dirichlet conditions
\begin{eqnarray}
R(\lambda,\rho_{max})&=&R_{max},\\
V(\lambda,\rho_{max})&=&0.
\end{eqnarray}
\noindent
The condition about the existence of the throat is imposed as a Newman condition
\begin{equation}
\partial_{\rho}R\vert_{\rho=0}=0.
\end{equation}

The asymptotic flatness imposes that $R \thicksim \mid \rho\vert + \mathcal{O}\left( l_{1} \ln \left(\mid \rho\vert/l_{2} \right) \right)$. Then, using equation (\ref{floweq2}), we are able to show that $V \thicksim \mathcal{O} \left( 1/\rho^5 \right)$. It is our immediate goal to find an initial metric, and a $V^i$ vector, that satisfies all the latter requirements. Note that the $V^{i}$ vector does not evolve under the flow, it is only restricted by equation \eqref{floweq2}, which justifies a certain freedom when choosing $V^{i}$. This freedom is the realization of the  invariance under diffeomorphisms of the flow. We have to find a specific $V^{i}$ for our initial metric, therefore our initial boundary value problem is well defined.

The time symmetric foliation of the well known Morris-Thorne wormhole \cite{Morris:1988cz} is
\begin{equation}\label{metric3}
ds^2=\frac{1}{g(r)}dr^2+r^2(d \theta^2+\sin(\theta)^2d\phi^2); \,\,\,\,\,\,\,\,g(r)=1-\frac{b(r)}{r},
\end{equation}

Then, if the shape function $b(r)$ is written as 
\begin{equation}
b(r)=\frac{b_{o}^{\delta}}{r^{\delta-1}}.
\end{equation}
\noindent
where $b_{o}$ has units of length and $\delta>-1/2$; the metric (\ref{metric3}) becomes
\begin{equation}\label{init:1}
ds^2=\frac{dr^2}{1-\left(\frac{b_{o}}{r}\right)^{\delta}}+r^2(d\theta^2+\sin^2(\theta)d\phi^2).
\end{equation}

In order to write this wormhole metric in the form (\ref{metric1}), we set
\begin{eqnarray}
r&=&R(0,\rho);\\
\partial_{\rho}R(0,\rho)&=&{\sqrt{g(R(0,\rho))}}.
\end{eqnarray}

Then, it is direct to obtain
\begin{equation}
\frac{\partial R(0,\rho)}{\partial \rho}=\sqrt{1-\left(\frac{b_{o}}{R(0,\rho)}\right)^{\delta}}.
\end{equation}

The solution to the previous differential equation is given in terms of the hypergeometric function ${}_{2}F_{1}(a,b,c,z)$ as
\begin{equation}\label{hyper}
{}_{2}F_{1}\left(\frac{1}{2},-\frac{1}{\delta},\frac{1-\delta}{\delta},\left(\frac{b_{o}}{R(0,\rho)}\right)^{\delta}\right)R(0,\rho)= \lvert \rho \rvert+{}_{2}F_{1}\left(\frac{1}{2},-\frac{1}{\delta},\frac{1-\delta}{\delta},\left(b_{o}\right)^{\delta}\right),
\end{equation}
\noindent
where we have used the condition $R(0,0)=1$ for determining the integration constant\footnote{It is known that this flow does not develop horizons, therefore if they are not present at the beginning of the flow they will not emerge as a result of the evolution \cite{LassoAndino:2018zhb}.}. From the previous equation, we see that $R(0,\rho)$ is dependent on $\delta$. When $\delta=2$, equation (\ref{hyper}) reduces to $R^2(0,\rho)=\rho^2+b_{o}^2$. In Fig. \ref{condinit} we have plotted the initial condition $R(\rho,0)$, with $\delta=1.3$. In the right panel, we show a revolution plot of (\ref{hyper}). We can see that there is a zone of saddle points (the throat). In compact 3-dimensional manifolds it is expected that the singularity appears at the throat. As we focus in asymptotically-flat spaces, we expect that asymptotic flatness forces to slow down the appearance of singularities, although at the end they will always arise.  
\begin{figure}[!ht]
    \centering
    \includegraphics[width=0.4\linewidth]{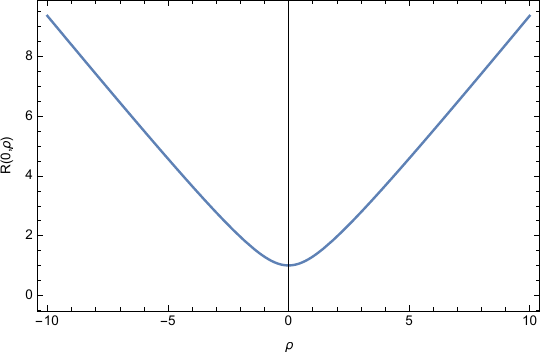}
    \includegraphics[width=0.3\linewidth]{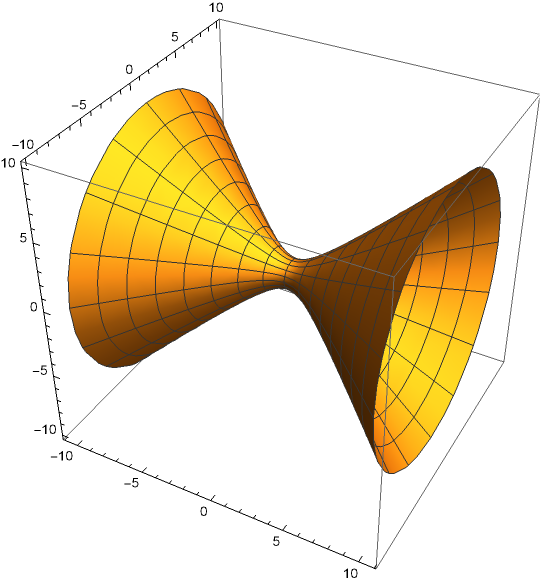}
    \caption{Initial condition $R(\rho,0)$, for $\delta=1.3$.}
    \label{condinit}
\end{figure}

\section{Wormhole evolution}\label{four}

Contrary to what happens with the Ricci flow, the evolution of any Riemannian metric under RG-2 flow has a restriction over the value of the scalar curvature. In other words, the system (\ref{floweq1})-(\ref{floweq2}) is going to be a parabolic system if and only if $1+ \alpha R_s/4>0$, where $R_s$ is the scalar curvature of the metric. When solving the system numerically, we have to ensure that the restriction is satisfied. \\
Moreover, we will normalize using the length scale $b_o$. Therefore, at the end, the equations will be independent of $b_{o}$. Thus,

\begin{equation}
 {\hat R} = R/b_o; \;\;\; {\hat \rho} = \rho/b_o; \;\;\; {\hat t} = t/b_o^2. 
\end{equation}

In what follows, we will omit the {\it hat}-notation. The RG-2 flow system has been solved in a periodic spatial domain $D = [-L,L]$, which is discretized with $N_\rho$ grid points in $D$. For the numerical evolution, we implement a pseudo-spectral method. The advances in $\lambda$ will be done with finite differences. However, as we want to study initial conditions involving high gradients in $D$, spectral methods are employed to computing first- and second-order derivatives respect to $\rho$. Computational cost of this method is similar to that of the pure finite-differences approach (not shown here), due to the domains size. The step $\Delta \lambda$ has been chosen in such a way that the Courant-Friedrichs-Lewy condition is always satisfied and keeps numerical noise low in order to study the RG-2 perturbative regime.  For all of our cases of study, the dimensionless function $V$ has to satisfy the conditions explained in Section \ref{initialc}. Thus, we use 
\begin{equation}
 V = \left( \frac{\rho - L}{10L} \right)^2
	 \exp \left[ -\frac{(\rho-\mu)^2}{2L} \right],
 \label{V3}
\end{equation}
\noindent
with $\mu = - 3 \pi + \sqrt{8 + 9 \pi^2}/2$. In the system (\ref{floweq1})-(\ref{floweq2}), the value $\alpha$ has to be taken in such a way that the influence of the higher curvature term becomes important. Remembering that the RG-2 flow comes from a perturbative expansion\footnote{The renormalization group flow of the non-linear $\sigma$-model has to be calculated perturbatively. The linear term in $\alpha$ is the Ricci flow, and the quadratic term in $\alpha$ corresponds to the RG-2 flow contribution}, we will work with $\alpha$ values going from the linear up to the quasi-nonlinear regime. \\
As we already pointed out, the family of metrics (\ref{hyper}) is parametrized by $\delta$. When $\delta=1$, we obtain the spatial sections of the Schwarzschild wormhole. When $\delta=2$, we get the Morris-Throne wormhole. We will also explore what happens during the evolution of metrics defined by different values of $\delta$.

Note that in our numerical set up the asymptotic flatness is numerically approximated. The compact domain $D$ is finite and therefore we have to check that our metric is mostly flat at the boundaries. We set the initial metric at the boundaries by setting $R_{max}$ at  $\rho_{max}$. Numerically, asymptotic  flatness will depend on how big is the size of $D$. For a large enough $D$ the curvature near the throat decreases to the point that it can be considered to have disappear.

\subsection{The Ricci flow}\label{sub:delta}
It has been suggested that, under Ricci-flow evolution, the wormhole pinches off (or expand) at the throat forever, depending on the values of initial data parameters \cite{Husain:2008rg}. Here, we present a numerical study of the evolution of the initial condition $R(0,\rho)$ --given in (\ref{hyper})-- under the Ricci flow. In Table \ref{table1}, we present the setup for three different runs. These parameters were chosen in order to test the evolution at different characteristic scales. 
\begin{table}[h!]
 \centering
 \begin{tabular}{ |l|c|c| }
 \hline
   & $L$ & $N_\rho$ \\
 \hline
 RUN 1 & $2 \pi$ & 64 \\ 
 RUN 2 & $4 \pi$ & 256 \\
 RUN 3 & $10 \pi$ & 512 \\
 \hline 
\end{tabular}
\caption{Simulations setup.}
\label{table1}
\end{table}
\begin{figure}[h!]
    \centering
    \includegraphics[width=0.3\linewidth]{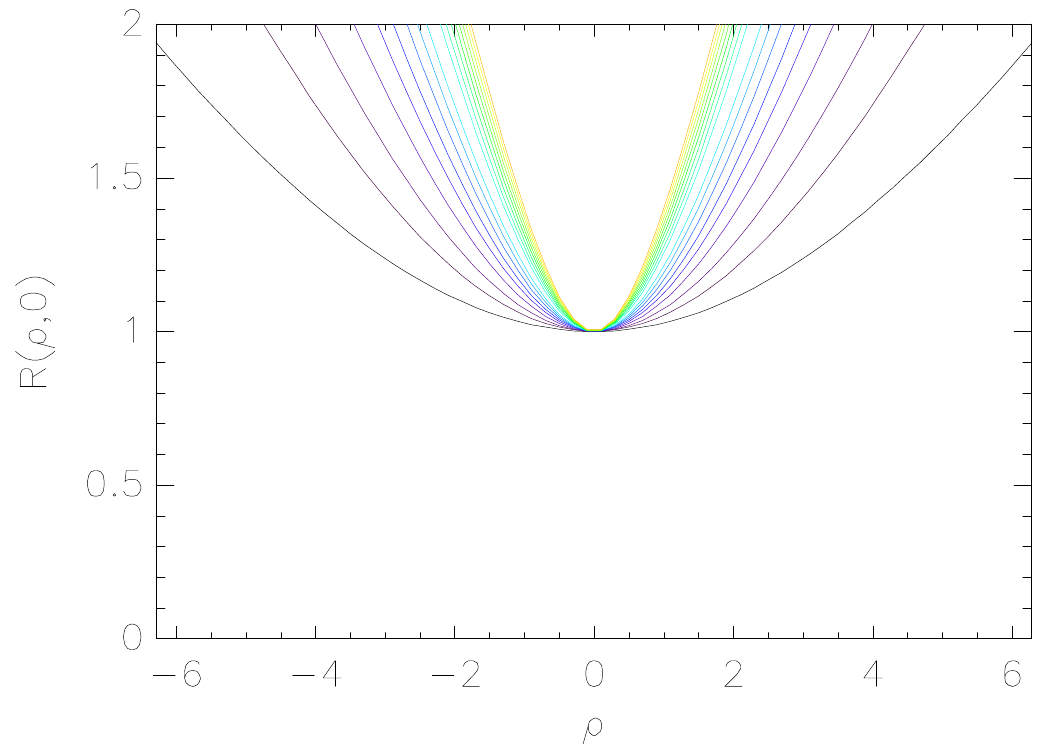}
    \includegraphics[width=0.3\linewidth]{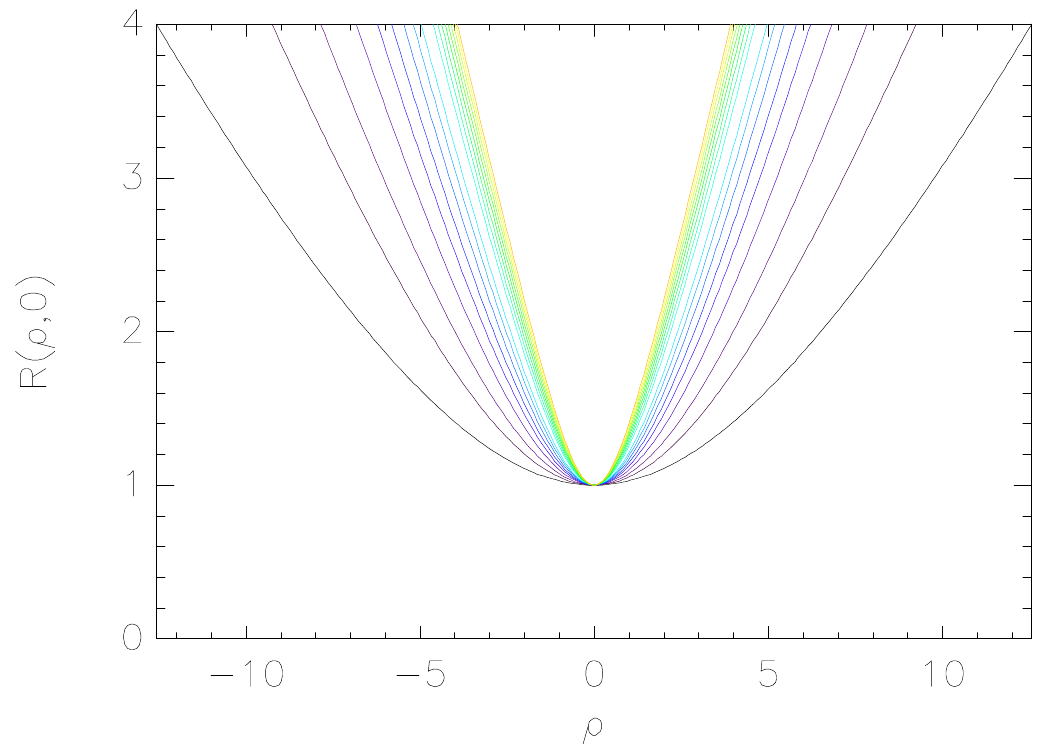}
    \includegraphics[width=0.3\linewidth]{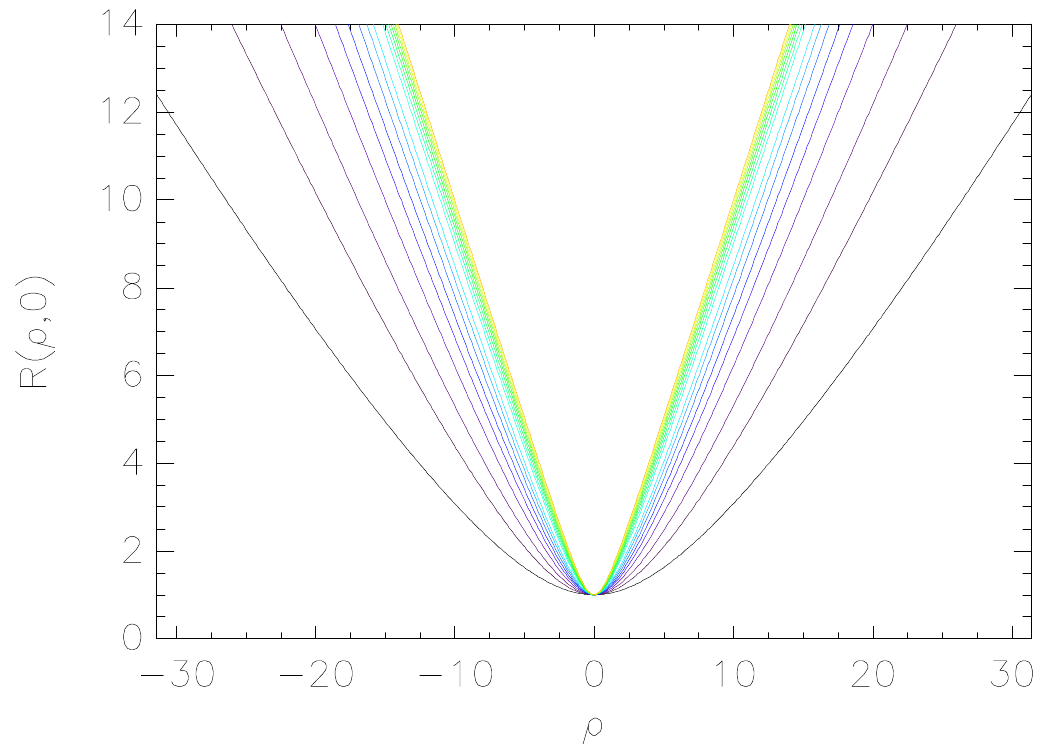}
    \caption{Initial conditions $R(\rho,0)$ chosen for RUN 1 (left panel), RUN 2 (middle panel), and RUN 3 (right panel). Each profile is a solution of equation (\ref{hyper}), for different values of $\delta$. The scale of colors departs from black ($\delta = 0.1$), and arrives to orange ($\delta = 1.9$).}
    \label{condinit}
\end{figure}

\begin{figure}[h!]
    \centering
    \includegraphics[width=0.3\linewidth]{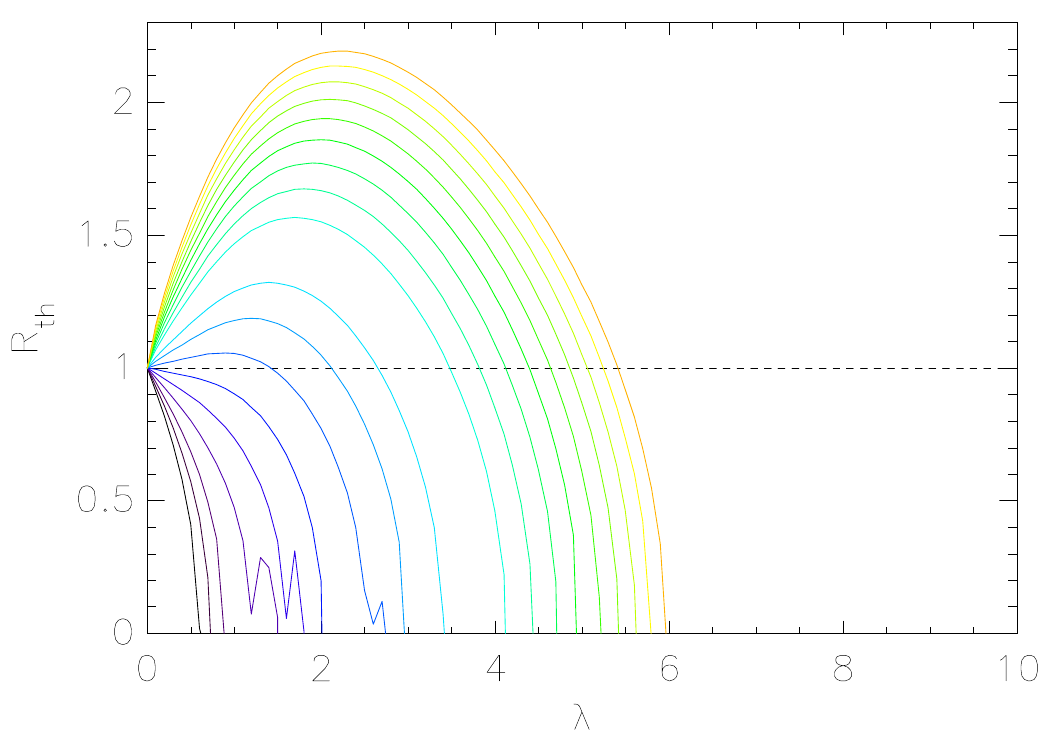}
    \includegraphics[width=0.3\linewidth]{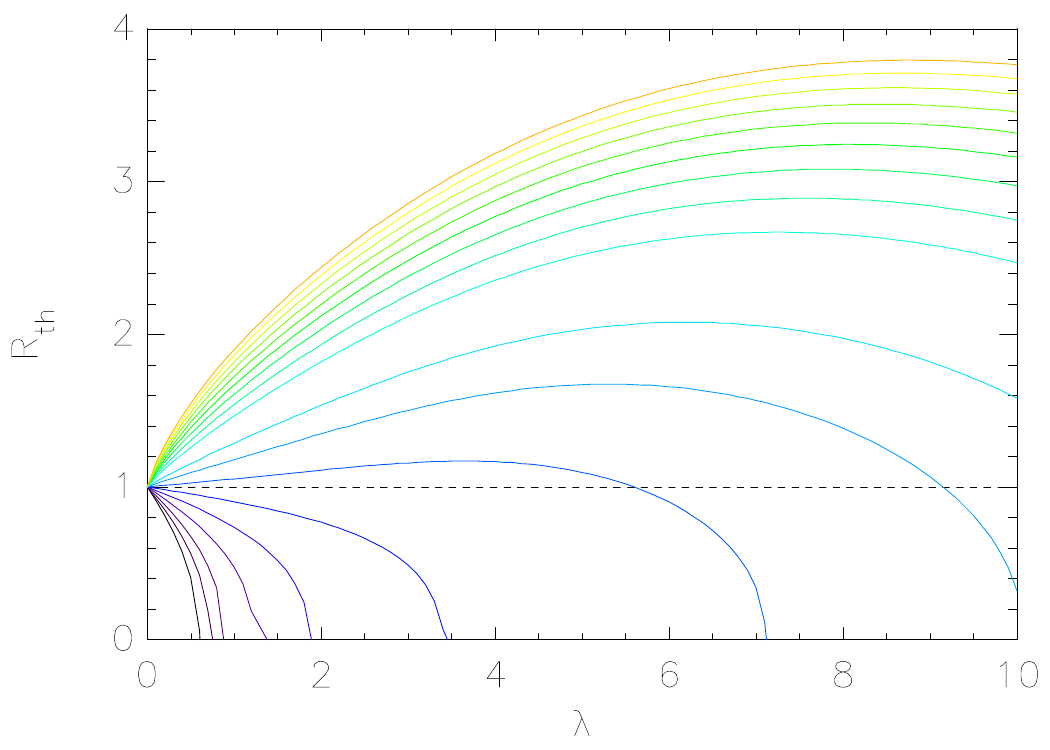}
    \includegraphics[width=0.3\linewidth]{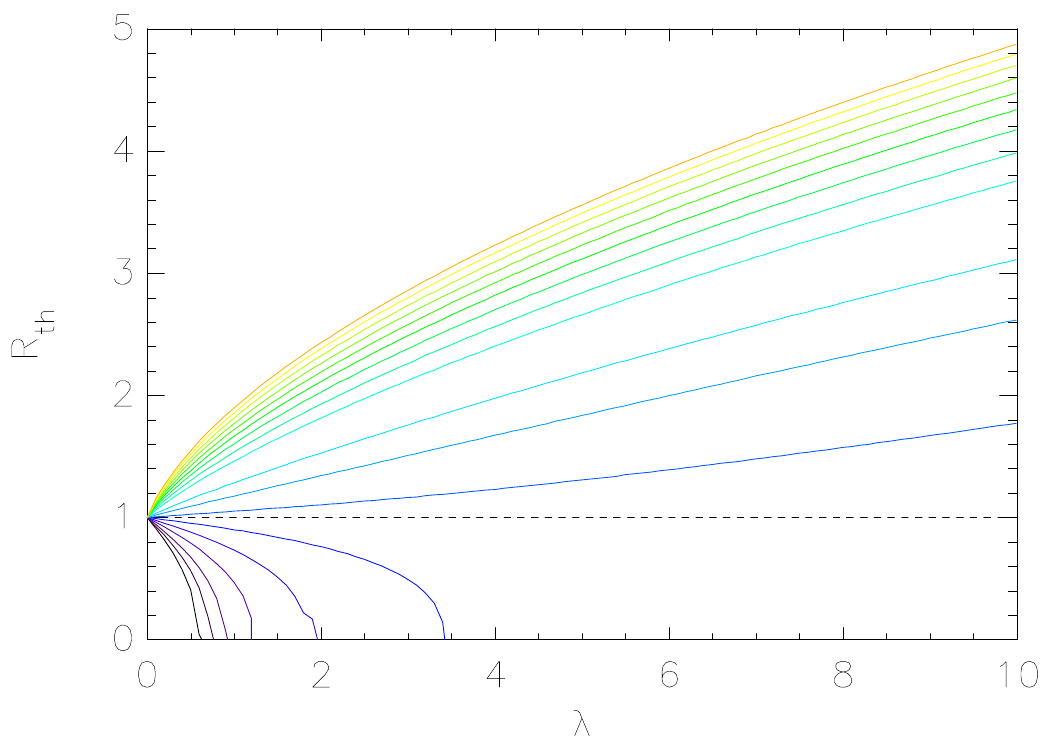}
    \caption{Evolution of $R_{th}$ in function of $\lambda$, using different values of $\delta$ (see the text) in RUN 1 (left panel), RUN 2 (middle panel), and RUN 3 (right panel). The scale of colors departs from black ($\delta = 0.1$), and arrives to orange ($\delta = 1.9$).}
    \label{throat}
\end{figure}

Figure \ref{condinit} shows the initial conditions $R(\rho,0)$ used for RUN 1 (left panel), RUN 2 (middle panel), and RUN 3 (right panel). Each initial condition is a numerical solution of equation (\ref{hyper}), for different values of $\delta$. In particular, We take 
\begin{equation}
\delta = \{ 0.1, 0.2, 0.3, 0.4, 0.5, 0.6, 0.7, 0.8, 0.9, 1.1, 1.2, 1.253, 1.3, 1.4, 1.5, 1.6, 1,7, 1.8, 1.9 \}\nonumber.
\end{equation}
In this Figure, the scale of colors departs from black, for $\delta = 0.1$ (lower line), and arrives to orange, for $\delta = 1.9$ (higher line). We will keep the same scale of colors throughout this Section. Note that for all of the initial conditions $R(0,0) = 1$.\\
The set of initial conditions, for RUN 1, RUN 2, and RUN 3, are evolved with the Ricci flow, {\it i.e.}, setting $\alpha = 0$ in the system (\ref{floweq1})-(\ref{floweq2}). In particular, we follow the minimum point of the throat $R_{th} \equiv R(0,\lambda)$, as a function of $\lambda$. The three panels of Figure \ref{throat} report the evolution of $R_{th}$ for RUN 1 (left panel), RUN 2 (middle panel), and RUN 3 (right panel), and for the different values of $\delta$, previously chosen. We note that for early stages of the simulation, $\lambda \lesssim 0.5$, no difference can be seen among the three runs, for the corresponding values of $\delta$. After this initial stage, and for values of $\delta > 0.3$, the characteristic length (dimension of the simulation box) affects the evolution of $R_{th}$, taking this value quicker to zero when $D$ decreases. In fact, for the latter range of $\delta$ values, $R_{th}$ remains positive for longer periods of $\lambda$ depending on $D$. We could name $\delta \approx 0.7$ as a ``critical'' value. However, it is clear, when comparing the three runs, that this behavior is apparent. As it is known from the mathematical literature, these kind of intrinsic flows smooth-out  the geometries when diffusing curvature from high curvature points to the lower ones. In our case, the asymptotic flatness of the metric plays an important role. Even though the asymptotic flatness tend to slow down the convergence to a singularity, at the end the flow wins and the singularity appears. Moreover, we note that the increase (or reduction) of the throat is not unlimited. In the cases of initial conditions with $\delta < 0.7$, the throat decreases from the beginning of each one of our simulations. However, in the cases of initial conditions with $\delta \geq 0.7$, the increment of $R_{th}$ is temporal. Even for RUN 3, where no decrease exist in the chosen interval of $\lambda$ (when $\delta \geq 0.7$), we could presume that $R_{th}$ will eventually decrease if a longer interval is chosen.

\subsection{From Ricci flow to RG-2 flow}
\label{sub:riccitorg2}

\begin{figure}[h]
    \centering
    \includegraphics[width=0.4\linewidth]{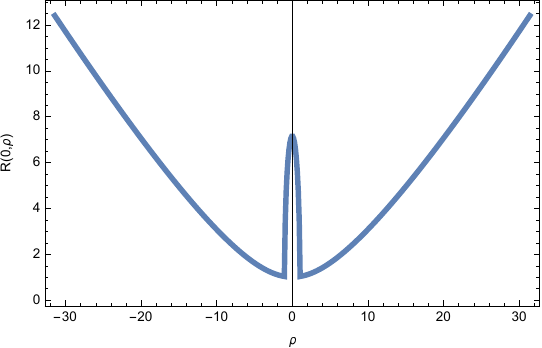}
    \includegraphics[width=0.4\linewidth]{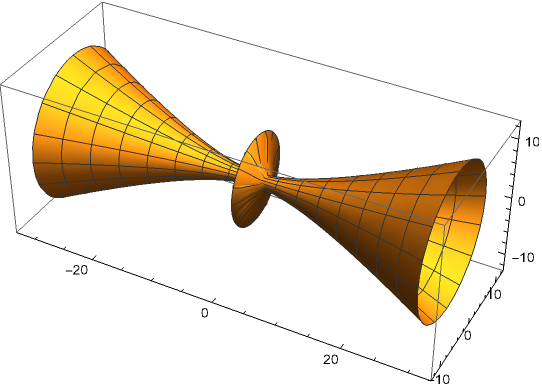}
    \caption{Initial condition $R(\rho,0)$, for $\delta=1.3$, with an embedded ellipse with minor axis $a=2$ and major axis $b=6$.}
    \label{condinit2}
\end{figure}

Now, we study the evolution of the wormhole metric (\ref{metric1}) under the RG-2 flow, {\it i.e.}, we take $\alpha \neq 0$ when solving the system (\ref{floweq1})-(\ref{floweq2}). We note that the idea of perturbing the Ricci flow equations leads to a better behavior when dealing with higher curvatures. Because of that, we maintain the $V^{i}$ given in \eqref{V3}. We have solved numerically the RG-2 flow equations for different values of $\alpha$, however, as expected, there are few differences respect to the evolution under the Ricci flow. Since the parameter $\alpha$ comes from a perturbative expansion\footnote{The parameter $\alpha$ is related with the length of a string in the non-linear sigma model.}, it has to be small. Therefore, in order to see the effect of the term $R_{a\alpha\beta\gamma}R_{b}^{\,\,\,\alpha\beta\gamma}$ we must built an initial wormhole metric such that the Riemann curvature tensor is big at some points. We focus in wormholes with strong curvatures. We take as initial metric the wormhole function (\ref{hyper}), where we have have dropped the throat section, and embedded a half of an ellipse with minor axis $a=2$ and major axis $b=6$. This initial condition $R(\rho, 0)$ is presented in Fig. \ref{condinit2}.

\begin{figure}[!ht]
    \centering
    \includegraphics[width=0.3\linewidth]{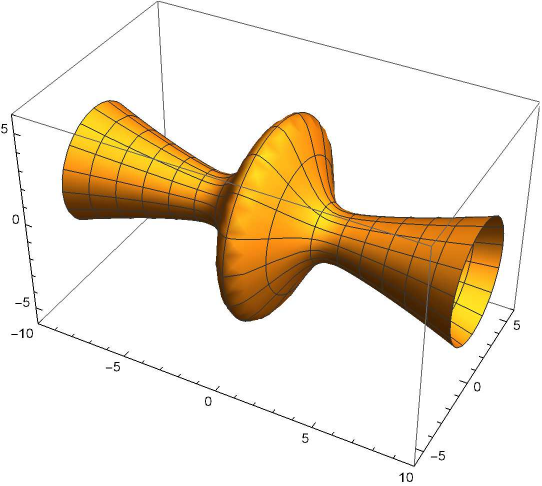}
    \includegraphics[width=0.3\linewidth]{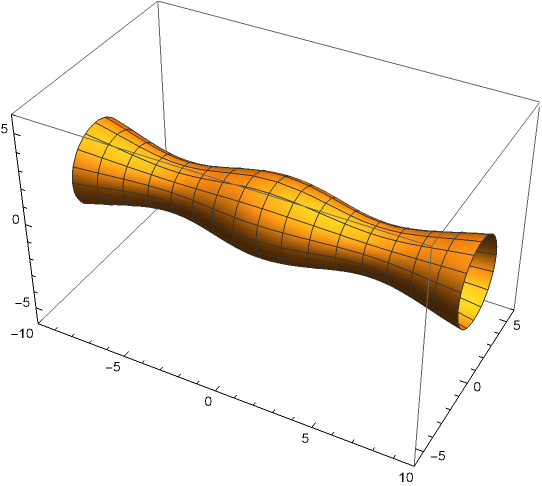}
    \includegraphics[width=0.3\linewidth]{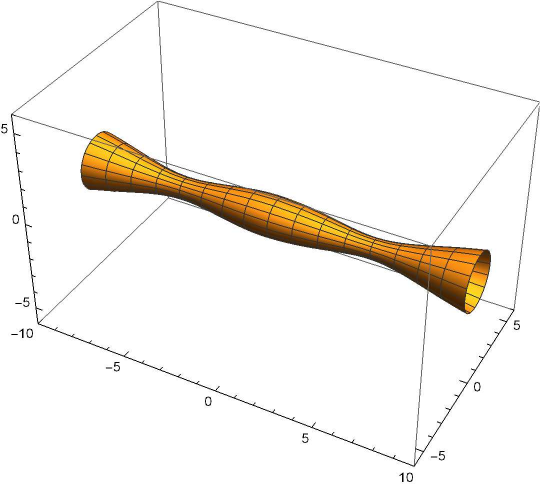}
    \caption{RG-2 evolution of the initial metric depicted in Figure \ref{condinit2}, at $\lambda =0.5$ (left panel), $\lambda = 5$ (middle panel), and $\lambda = 8$ (right panel).}
    \label{fig:highc}
    \end{figure}

\begin{figure}[!ht]
    \centering
    \includegraphics[width=0.4\linewidth]{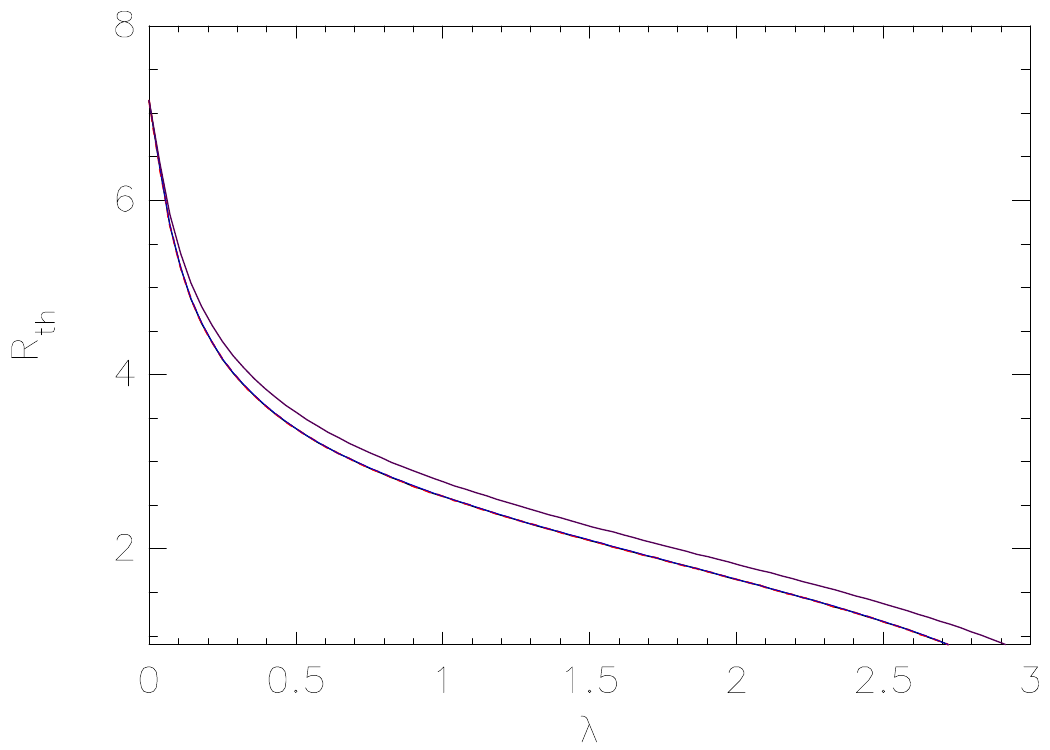}
    \caption{Evolution of $R_{th}$ in the system (\ref{floweq1})-(\ref{floweq2}), with $\alpha = 0$ (red line), $\alpha = 0.001$ (black line), $\alpha = 0.01$ (blue line), and $\alpha = 0.15$ (purple line). Only the last curve is not overlapped with the rest of values of $\alpha$.}
    \label{fig:rthcurv}
\end{figure}

\begin{figure}[!ht]
    \centering
    \includegraphics[width=0.45\linewidth]{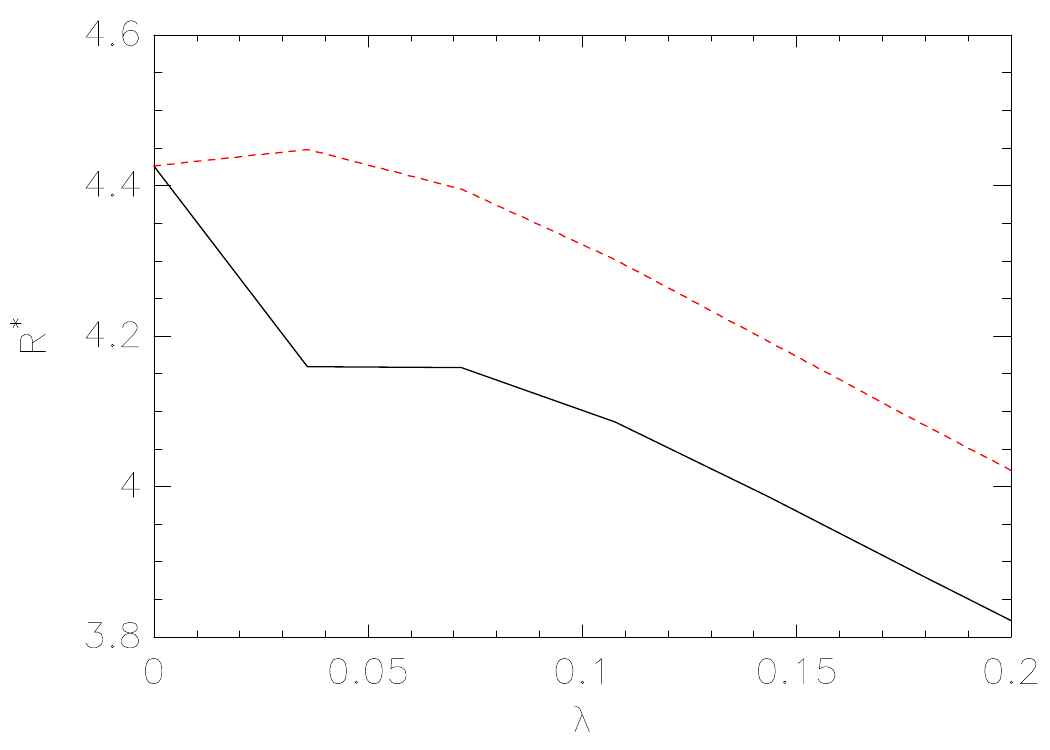}
    \includegraphics[width=0.45\linewidth]{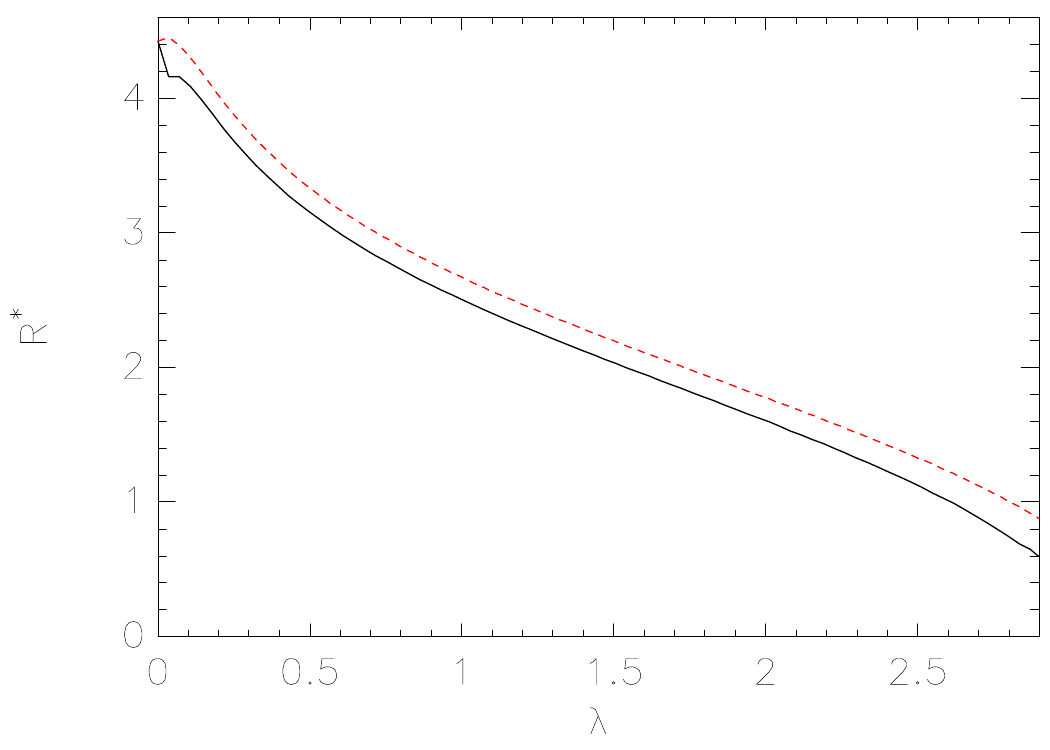}
    \caption{$\lambda$-evolution of $R^*(\lambda)$ (see text). The left panel emphasizes the initial instants of the evolution of $R^*$ with the system (\ref{floweq1})-(\ref{floweq2}), and with $\alpha = 0$ (black-solid line), and $\alpha = 0.15$ (red-dashed line).}
    \label{fig:rasterisk}
\end{figure}

We follow the evolution of this initial condition, with different values of $\alpha$. We see in Figure \ref{fig:highc} three evolution snapshots --at $\lambda = 0.5$ (left panel), $\lambda = 5$ (middle panel), and $\lambda = 8$ (right panel)-- of the initial condition depicted in Figure \ref{condinit2}, with $\alpha = 0.15$. Moreover, in Figure  \ref{fig:rthcurv}, we show the evolution of $R_{th}$ with $\alpha = 0$ (red line), compared with $\alpha = 0.001$ (black line), $\alpha = 0.01$ (blue line), and $\alpha = 0.15$ (purple line). Note that the two first curves are overlapped, and that for $\alpha = 0.15$, and $\lambda \gtrsim 0.1$, $R_{th}$ decreases slowly respect to the other cases. As it was previously noted in Subsection \ref{initialc}, we expect that this behavior becomes more notorious in regions where the curvature is higher. For example, the panels of Figure \ref{fig:rasterisk} show the evolution of $R^*(\lambda) \equiv R(\rho = 0.8,\lambda)$, for $\alpha = 0$ (black-solid line), and $\alpha = 0.15$ (red-dashed line). It is clear that for early stages of the simulation $\lambda \lesssim 0.1$ (left panel), the value of $\alpha$ is quite important differentiating the future of $R^*$. After this early stage (right panel), the evolution of $R^*$ is similar for both values of $\alpha$.\\
We can see that, depending on the value of $\alpha$, we need to make evolve a metric with a very high curvature in order to see a difference between both flows. The initial condition depicted in Fig. \ref{condinit2} has two throats (at each side of the ellipse) and can still be called a wormhole. But, since it was constructed for satisfying curvature requirements, it might not be a solution of the Einstein equations. The evolution after a long time under the RG-2 flow is very similar to the Ricci flow case. We note that when the Ricci flow develops a singularity the RG-2 flow is able to continue a little bit further.

\subsection{Entropy and Energy}
\begin{figure}[h]
    \centering
    \includegraphics[width=0.45\linewidth]{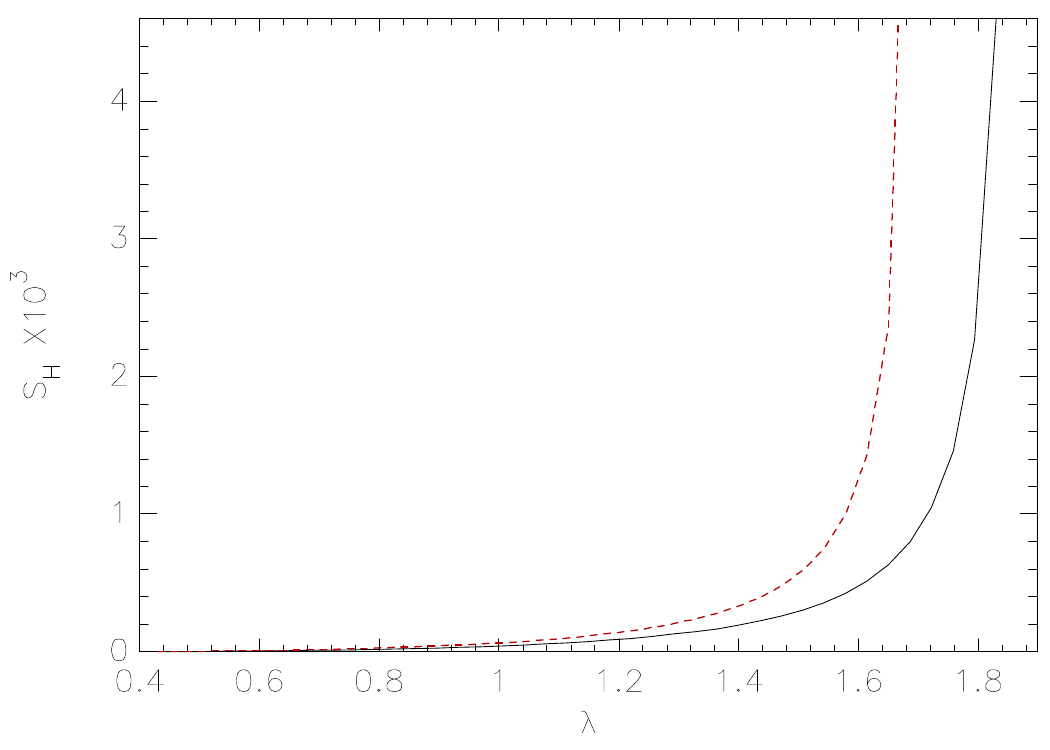}
    \includegraphics[width=0.45\linewidth]{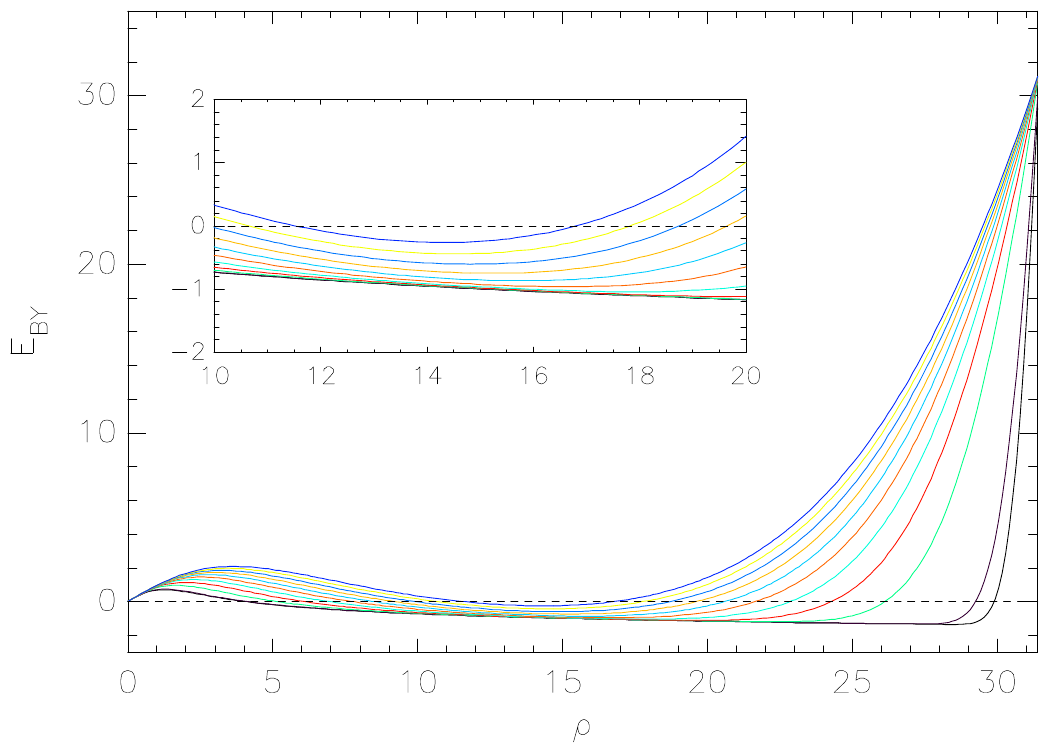}
    \caption{Entropy ($S_{H}$) vs $\lambda$ in the left panel. The red line is the entropy $S_{H}$ when evolved by the RG-2 flow, and the black line is the evolution under the Ricci flow. In the right panel, we have plotted the Brown York energy ($E_{BY}$) depending on $\rho$. Each curve represents $E_{BY}$ at different $\lambda$'s.}
\label{evse}
\end{figure}

When a manifold evolves under a intrinsic flow, we can calculate the evolution of certain quantities, that somehow let us characterize what is happening with the whole geometry. It would be very good if we are able to find a quantity that tell us if the flow has stable points, it means that we will be able to know, at least numerically, if we have an integrable /renormalizable theory.\\
Here, we study two quantities: the Brown-York energy $E_{BY}$, which is a quasi-local quantity, and the Hamilton's entropy $S_{H}$, which is a global quantity. \\
Due to the fact that we are evolving a time-symmetric foliation of a Lorentzian spacetime, we should use a quantity defined in a Riemannian metric but giving information about the Lorentzian metric. In order to do so, we will use the Brown-York energy, which for the metric (\ref{metric1}) is given by
\begin{equation}\label{BYener}
E_{BY}=\rho\left(1-\frac{\rho\, \partial_{\rho}R}{R}\right).
\end{equation}
The Brown-York quasi-local energy measures the gravitational energy contained in a spacelike surface defined by $\rho=const.$ When $\rho\rightarrow  \infty$ we get the total energy contained in the whole space. When $R(\rho)=\rho$ (Minkowski space) then $E_{BY}=0$. It is well known that $E_{BY}$ depends only on the spatial curvature of the metric, therefore when our flow evolves the spatial metric (\ref{metric1}) the energy $E_{BY}$ is going to change also.\\
In the right panel of Fig. \ref{evse}, we show the evolution of the Brown-York energy with $\delta=1.0$. Every curve represents the $E_{BY}$ at different instants of evolution. We can see that the energy contained in a surface of constant radius $\rho$ goes through some stages depending on how big is $\rho$. The initial point is zero (because $\rho=0$) for all curves. Then, we can divide the domain of the function in three zones. The zone from zero until the first root of the function, where the $E_{BY}$ becomes positive; it starts growing, but later descends and becomes negative, where another zone starts. This is the longest zone at the beginning of the evolution but it shrinks through evolution. In General Relativity, a zone where the $E_{BY}$ is negative has a known physical interpretation. This is a zone where gravity is attractive in the Newtonian sense\footnote{In the Schwarzschild metric, the zone outside the horizon has negative $E_{BY}.$}. The third zone is when $E_{BY}$ becomes positive again and starts growing with $\rho$. This behavior is expected, the energy contained in a surface $\rho=const$ is going to change because the metric is changing. When $\delta=0.1$ (not shown here) the energy $E_{BY}$ is, although increasing in certain zones and decreasing in others, positive everywhere for all $\rho$ values.
There are certain quantities that are monotonous under the RG-2 flow. In order to check that our numerical solution is working correctly, we evolve a quantity, known as monotonous under the RG-2 flow. We take $S_{H}$ and compute its evolution when the metric (\ref{metric1}) evolves under the RG-2 flow. This Hamilton's entropy is inspired in the classical Boltzman definition, but it uses curvature. In \cite{LassoAndino:2019lsa}, it was shown that the formula for Hamilton's entropy can also be used for asymptotically flat spaces, and for the flow without normalization. For the normalized flow case, see \cite{Branding:2015}. Therefore, we can compute numerically the entropy of the wormhole surface (see appendix \ref{apenn}).\\
\pagebreak
The entropy of the surface of revolution will grow with $\lambda$. The enclosed volume shrinks and asymptotic flatness is maintained. If the surface would be compact and the constancy of the volume is enforced, the surface will develop singularities with a similar geometry of those of the Ricci flow\footnote{In the Ricci flow  with surgery used in the proof of the Thurston geometrization theorem, Perelman was able to remove the singularities and let the flow continue towards one of the 8 Thurston geometries. The requirement was the constancy of the volume form multiplied by an exponential factor.}. By definition, the Hamilton's entropy for the RG-2 flow is 
\begin{equation}
S_{H}=\int_{\mathcal{M}}\left(R_{s}\,\log (R_{s})+\frac{\alpha}{2}R_{s}^2\right)d\mu,
\end{equation}
where $d\mu$ is the volume element, and $R_{s}$ is the scalar curvature of the manifold. This entropy is monotonous under the RG-2 flow\footnote{Depending on the sign chosen this entropy will be increasing or decreasing.}. In Fig. \ref{evse} , left panel, we can clearly see the monotonicity of $S_{H}$, this property is an indirect confirmation that our numerical evolution is working well. This entropy is indeed monotonous under both flows, the Ricci flow (black line) and under the RG-2 flow (red line). We have not fixed the volume and we have enforced asymptotic flatness. In our case, this entropy is growing with lambda and at the end of the interval it diverges. We expect that when the number of loops grows the entropy $S_{H}$ is going to be still increasing with the flow.

\section{Discussion and Conclusions}\label{five}

We have evolved different wormhole geometries under the Ricci flow, and we have perturbed the flow in order to analyze higher curvature terms. Thus, the RG-2 flow is considered as a perturbative expansion to the Ricci flow, the numerical solutions deviate from the Ricci flow ones controlled by the RG-2 term. In all cases in our study, we kept the numerical error low enough not to interfere with the perturbative terms. When curvatures are small there is practically no difference between both flows. However, when curvatures grow we can clearly see differences. Although both flows develop singularities, the RG-2 flow is able to handle high curvatures better. We have used a general definition of wormhole, although we have studied only the asymptotically flat case. Moreover, we have studied the evolution of wormhole-like metrics that are not solutions of Einstein Equations. We have found that the wormhole pinches-off no matter what kind of initial condition we start with, sooner o later we will have a singularity.
An interesting result is the fact that the apparent criticality described in \cite{Husain:2008rg} appears only in certain zone. If we let the flow evolve further, we can see that all the studied wormholes pinch-off.  This is an important result, we are showing numerically that there is no integrability/renormalizability for the non-linear $\sigma$-model when its target space corresponds with the spatial sections of the Morris-Thorne wormhole with $\delta=1.259$. We confirmed the result by evolving the same geometry with the RG-2 flow, which deals better with higher curvature zones. Due to the fact that higher loops flows will add more higher curvature terms, and the behavior for low curvatures is the same for both flows, we can safely conjecture that for the type of wormhole solutions studied here, the singularity formation will take place at all loops and for large $\lambda$'s.
Another interesting fact is that in our evolution the entropy grows, enforcing the reduction of the enclosed volume. If the wormhole has a throat, which introduces the curvature in the wormhole, it starts expanding but the asymptotic flatness forces the throat to retreat. Thus, in order to maintain asymptotic flatness, the surfaces shrinks and the enclosed volume will reduce.\\
The Hamilton's entropy computed for the surface is a good indicator of what is going on through the evolution. The entropy is going to be monotonous, and according to our estimation, it will be an increasing function of $\lambda$. When the enclosed volume reduces, $S_H$ grows. We note that the geometric entropy has been used for finding bounds on curvature evolution, but a physical interpretation is still missing. Recently, the interest for understanding this (and other kind of entropies) have increased. In \cite{Kehagias:2019akr}, the authors studied the entropy functionals of a gradient flow (the Ricci flow). They concluded that these entropies provide a good definition for the distance, in the context of the infinite distance conjecture in the swampland approach. The behavior of these entropies has not been studied extensively, and since there are many candidates for a geometric entropy, we believe that it deserves more study. Until now, we have shown that the entropy is monotonous, even without restricting the volume to be constant. We should study what happens with different geometries before arriving to a physical interpretation.
It is known that compact surfaces of constant curvature shrink to a point when evolved with the Ricci flow \cite{Rflow:1}. A similar behavior was expected with RG-2 flow evolution. However, in our set up, we enforced asymptotic flatness and therefore the flow should either goes to flat space or develops a singularity. This suggests that an interesting problem could be the evolution of asymptotically AdS wormholes \cite{Maldacena:2017axo}. A family of these wormholes is interpreted as connector of entangled particles. Therefore, in this context, the development of singularities would mean disentanglement. We could check what happens with dual fields in the conformal theory when its bulk counterpart are evolved under a geometric flow.\\

\section*{Acknowledgments}

C.L.V was partially supported by EPN internal projects PII-DFIS-2019-01 and PII-DFIS-2019-04.

\appendix
\section{Hamilton's Entropy}\label{apenn}

Here, we make a brief review of the Hamilton's entropy and its properties. In order to find some bounds to the curvature evolution under Ricci flow, in \cite{Hamilton:1986} Hamilton defined a geometric entropy, in analogy to the thermal entropy\footnote{It is known that Boltzman have used as a definition of entropy $S_{B}=\rho ln \rho$ where$\rho$ was a density phase space. Later on, Gibbs generalized the Boltzman entropy to the form $S_{G}=\sum p_{i}ln p_{i}$, where $p_{i}$ are probabilities}. The Hamilton's entropy is defined for closed surfaces of positive curvature and is given by
\begin{equation}
S_{H}=\int_{M}\left(R_{s} \,\log (R_{s}) \right)dv,
\end{equation}
where $R_{s}$ is the scalar curvature of the manifold $\mathcal{M}$ and $dv$ is the volume form in two dimensions. $S_{H}$ is monotonous under normalized Ricci flow\footnote{The normalized Ricci flow on surfaces is defined as \begin{equation}
\frac{\partial g_{ij}}{\partial \lambda}=-2\left(R_{s}-\frac{\int_{\mathcal{M}}R_{s}dv}{\int_{\mathcal{M}} dv}\right)
\end{equation}} evolution (see Proposition 5.39 in \cite{Rflow:1}). When the scalar curvature is negative we have to use another, although very similar, formula for the entropy:
\begin{equation}\label{entropym}
S_{Hw}=\int_{M}(R_{s}-w) \,\log(\,R_{s}-w)dv,
\end{equation}
where 
\begin{equation}
w(r)=\frac{r}{1-\left(1-\frac{r}{w_{o}}\right)e^{rt}}.
\end{equation}

\begin{figure}[h]
    \centering
    \includegraphics[scale=0.6]{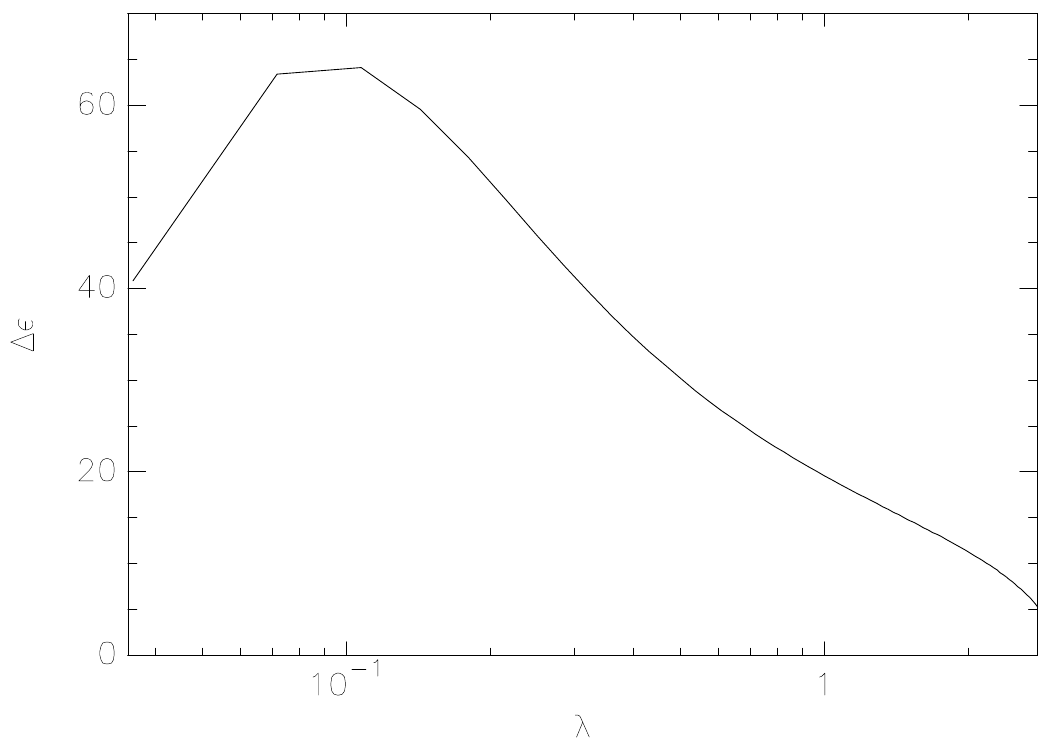}
    \caption{Energy difference between both flows. There is a maximum before the difference starts going to zero.}
    \label{enth}
\end{figure}

For the case of a negative curvature, $S_{Hw}$ is not necessarily monotonous, but is bounded from above (see Proposition 5.44 in \cite{Rflow:1}). These results can be extended for the RG-2 flow. In \cite{Branding:2015} there is a generalization of the Hamilton's entropy for the RG-2 flow, this new entropy is given by
\begin{equation}
S_{H}=\int_{M}\left(R_{s} \,\log(R_{s})+\frac{\alpha}{4}R_{s}\right) dv,
\end{equation}
where, as before, $R_{s}$ is the scalar curvature of the metric, and $dv$ is the volume form in two dimensions. The definition has been extended to asymptotically flat spaces, in \cite{LassoAndino:2019lsa}.
For our wormhole metric (\ref{metric1}) the scalar curvature is 
\begin{equation}
R_{s}=-\frac{2(1-(R')^2+2RR'')}{R^2}.
\end{equation}
Therefore, we can calculate the surface entropy at every time of the evolution.
Complementary, we take advantage of the fluid nature of the system (\ref{floweq1})-(\ref{floweq2}) for defining and computing an ``energy density'' $\epsilon \equiv R^{2}/2$, for each instant $\lambda$. This quantity would behave as its counterpart --the kinetic energy density in Navier-Stokes models.\\
At first sight the energy evolution of the Ricci flow is very similar to the RG-2 flow energy, but there are differences, specially when lambda grows. In Fig. \ref{enth} we have plotted the energy difference between both flows $\Delta \epsilon$ 

This difference shows that the RG-2 flow is dealing better with higher curvature. Moreover, it approaches differently the singularities. At the beginning  of the evolution the asymptotic flatness of the spacetime forces the wormhole to reduce the area. Since this energy is, in some way a measure of the curvature, we can see that each flow acts differently when smoothing curvatures, but at the end the result is the same, both flows tend to reduce the curvature and finally develop singularities.\\

\textbf{CRediT authorship contribution statement}\\
\textbf{O. Lasso Andino}: Conceptualization, Methodology, Formal Analysis, Writing - Original Draft. \textbf{C. L. V\'{a}sconez}: Methodology,Software,Writing - Review $\&$ Editing,Visualization.\\

\textbf{Declaration of competing interest}\\
The authors declare that they have no known competing financial interests or personal relationships that could have appeared to influence the work reported in this paper.\\

\textbf{Data availability}\\
No data was used for the research described in the article.

\end{document}